\documentclass[aps,pre,preprint,showpacs,eqsecnum,amssymb]{revtex4}
\usepackage{graphicx,color,psfrag,amsmath,bm}

\begin{document}
\title{
Heat fluctuations and fluctuation theorems
in the case of multiple reservoirs
}
\author{Hans C. Fogedby}
\email{fogedby@phys.au.dk}
\affiliation{Department of Physics and
Astronomy, University of
Aarhus, Ny Munkegade\\
8000 Aarhus C, Denmark\\}
\affiliation{Niels Bohr Institute, Blegdamsvej 17\\
2100 Copenhagen {\O}, Denmark}
\author{Alberto Imparato}
\email{imparato@phys.au.dk} \affiliation{Department of Physics and
Astronomy\\
University of Aarhus, Ny Munkegade\\
8000 Aarhus C, Denmark}
\begin{abstract}
We consider heat fluctuations and fluctuation theorems for
systems driven by multiple reservoirs. We establish a fundamental
symmetry obeyed by the joint probability distribution for the heat 
transfers and system coordinates. The symmetry leads to
a generalisation of the asymptotic fluctuation theorem for
large deviations at large times. As a result the presence of multiple
reservoirs influence the tails in the heat distribution.
The symmetry, moreover, allows for a simple derivation
of a recent exact fluctuation theorem valid at all times.
Including a time dependent work protocol we also present
a derivation of the integral fluctuation theorem.
\end{abstract}
\maketitle
\section{\label{intro}Introduction}
There is a current interest in the thermodynamics and statistical
mechanics of fluctuating systems in contact with heat reservoirs
and driven by external forces. The current focus stems from the
recent possibility of direct manipulation of nano systems and
biomolecules. These techniques permit direct experimental access
to the probability distribution functions for the work or for the
heat exchanged with the environment
\cite{Trepagnier04,Collin05,Seifert06a,Seifert06b,Wang02,
Imparato07,Imparato08,Ciliberto06,Ciliberto07,Ciliberto08,Ciliberto13,Ciliberto14,Koski13}. 
These methods have also yielded access to the experimental verification of the
recent fluctuation theorems which relate the probability of observing 
entropy-generated trajectories with that of observing
entropy-consuming trajectories \cite{Jarzynski97,Kurchan98,Gallavotti96,Crooks99,Crooks00,
Seifert05a,Seifert05b,Evans93,Evans94,Gallavotti95,Lebowitz99,Gaspard04,Imparato06,
vanZon03,vanZon04,vanZon03a,vanZon04a,Seifert05c,Imparato07,Esposito10a}.

In the present paper we address the issue of heat fluctuations and fluctuation theorems
in the case of a system coupled to multiple reservoirs. The case of a system driven by
two heat reservoirs, both in the case of one degree of freedom 
and in the case of many degrees of freedom, has been
discussed extensively, see e.g.  \cite{Fogedby11a,Fogedby12}, whereas
the case of multiple reservoirs seems to have received less attention.
In the case where a system is coupled to many reservoirs the heat flows
exhibit a more complicated pattern which will influence the heat distribution.

We characterise the heat transfers to the system by the vector quantity $\mathbf Q$ 
with components $\{Q_n\}$, referring to the n-th heat reservoir maintained at temperature
$T_n=1/\beta_n$; we set Boltzmann's constant $k_{\text{B}}=1$. We, moreover,  assume that the reservoirs  are independent.
Noting that the transfer of heat also influences the internal state
of the system characterised by the coordinate $x$, the central quantity is
the joint distribution $P(\mathbf Q,xt)$. For simplicity we consider a single degree of 
freedom $x$; the generalisation to many degrees of freedom is straightforward
and discussed at the end of the paper.

Assuming that the system is initially in the state $x_0$ at time $t_0$ and that 
the sampled heat vanishes at $t_0$, the joint distribution $P(\mathbf Q,xt,x_0t_0)$ 
describes the transition
of the system from the initial state $x_0$ at time $t_0$ to the final state $x$ at time $t$
in contact with multiple heat reservoirs transferring the heat $\mathbf Q$ to the system
during the time span $t-t_0$. We, moreover, assume that the system moves in the time 
independent potential $U(x)$. Hence, there are no external forces acting on the system and
the non equilibrium state is entirely driven by the heat transfers from the reservoirs.

It follows from the structure of Fokker-Planck equation that the  joint distribution  
$P(\mathbf Q,xt,x_0t_0)$ obeys the fundamental symmetry
\begin{eqnarray}
P(\mathbf Q,xt,x_0t_0)=\exp\Big(-\beta_k(U(x)-U(x_0))\Big)\exp\Big(-\sum_n\beta_{nk}Q_n\Big)P(-\mathbf Q,x_0t,xt_0),
\label{psym}
\end{eqnarray}
as discussed in Sec.~\ref{subsec:sym}.
Here we have singled out the k-th reservoir at temperature $T_k=1/\beta_k$
and, moreover, introduced the notation
\begin{eqnarray}
\beta_{nk}=\beta_n-\beta_k=1/T_n-1/T_k.
\label{not}
\end{eqnarray}
The symmetry  (\ref{psym}) which is valid at all times is quite general for systems driven stochastically
by heat reservoirs. The symmetry basically establishes a connection between the heat transfers
during a transition from $x_0$ to $x$ and minus the heat transfers during the reverse
transition from $x$ to $x_0$. The symmetry incorporates the general features of a
reservoir-driven non equilibrium transition. As will become clear, the symmetry implies 
both the asymptotic fluctuation theorem for the cumulant generating function valid
at long times, see e.g. \cite{Gallavotti95,Lebowitz99}, the integral fluctuation theorem 
\cite{Seifert05a}, and the exact fluctuation theorem proposed in \cite{Cuetara14} valid at all times.

In the present paper we discuss in detail the implications of the symmetry (\ref{psym}).
For the purpose of a simple discussion we introduce in Sec.~\ref{sys} the
model of a single over damped particle moving in a time independent potential 
and at the same time driven by
multiple heat reservoirs. This model allows for a simple derivation of the 
symmetry in (\ref{psym}). In Sec.~\ref{cum} we consider the asymptotic
long time regime and discuss the application of the symmetry to the 
cumulant generating function. Limiting the discussion
to the harmonic potential we consider the branch cut structure of the 
cumulant generating function and the implications for the tails in the
associated heat distribution. In Sec.~\ref{fluc} we discuss implications
of the symmetry valid at all times. We make contact with the trajectory
formulation and associated entropy considerations. We also present a 
derivations of the integral fluctuation theorem and the exact fluctuation 
theorem. In Appendix E we summarise the extension of the formalism
to the case of a time dependent potential modeling an external work protocol.
This extension allows us to demonstrate the integral fluctuation theorem 
in the general case. In Sec.~\ref{ham} we consider the case of a deterministic Hamiltonian
system driven by heat reservoirs and present the corresponding symmetry for 
the joint distribution. In Sec.~\ref{sum} we present a summery and a conclusion.
Technical matters are deferred to a series of appendices.
\section{\label{sys}System coupled to multiple heat reservoirs - one degree of freedom}
\subsection{General}
\begin{figure}
\includegraphics[width=1.0\hsize]{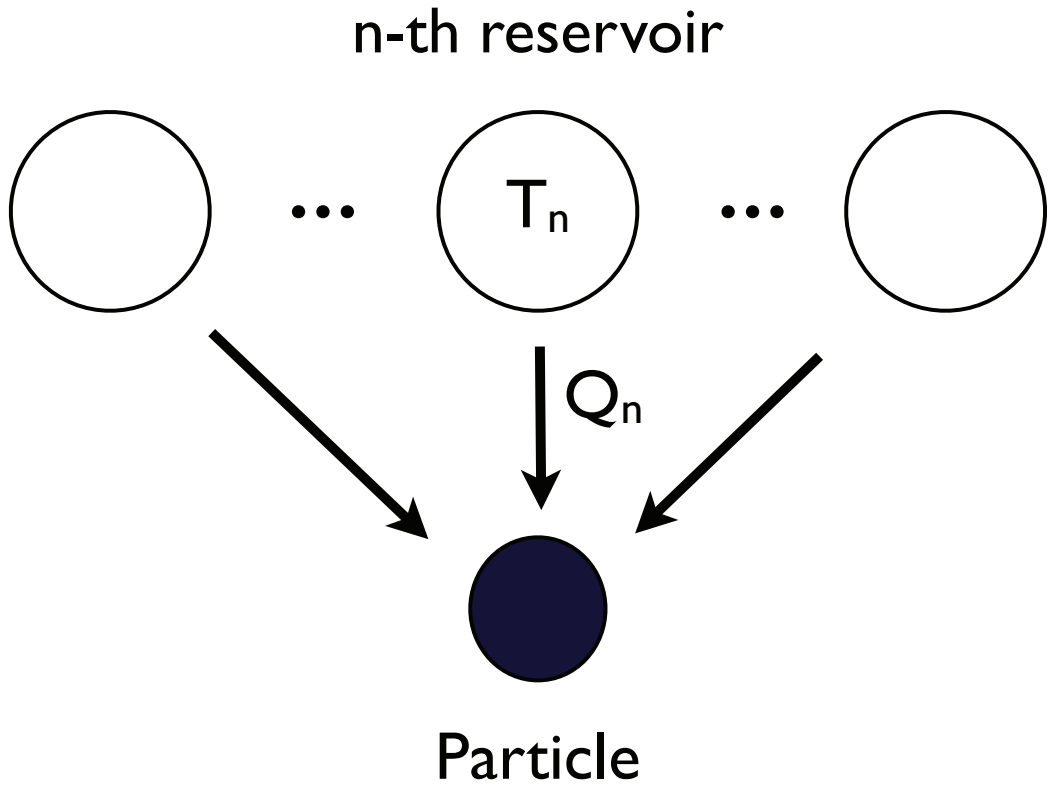}
\caption{We depict the configuration corresponding to a single degree of
freedom, a particle, coupled to several reservoirs. The temperature of the n-th
reservoir is maintained at $T_n$. The heat transferred to the particle from the
n-th reservoir is denoted $Q_n$.}
 \label{fig1}
\end{figure}

Here we consider for simplicity a single degree of freedom $x$ moving in the potential
$U(x)$ and at the same time driven by multiple heat reservoirs maintained
at temperatures $T_n=1/\beta_n$. We restrict our discussion to the autonomous case where the potential is
time independent. Associating the damping $\Gamma_n$ with the n-th reservoir, 
the corresponding Langevin equation has the form, see e.g. \cite{Imparato07,Seifert05a},
\begin{eqnarray}
\frac{dx(t)}{dt}=-\Gamma U'(x)+\xi(t),
\label{lan1}
\end{eqnarray}
where we have introduced the total damping, the total noise,
and the noise correlations, i.e.,
\begin{eqnarray}
&&\Gamma=\sum_n\Gamma_n,
\label{damp}
\\
&&\xi(t)=\sum_n\xi_n(t),
\label{noise}
\\
&&\langle\xi_n(t)\xi_m(t')\rangle=2\Gamma_nT_n\delta_{nm}\delta(t-t');
\label{corr}
\end{eqnarray}
here a prime denotes the spatial derivative $d/dx$.
Correspondingly, the heat flux from the n-th reservoir is given by
$F_n(x)U'(x)$, where $F_n(x)=-\Gamma_n U'(x)+\xi_n(t)$ is the force on the 
system originating from the n-th reservoir, i.e.,
\begin{eqnarray}
\frac{dQ_n(t)}{dt}=-\Gamma_n(U'(x))^2+U'(x)\xi_n(t).
\label{lan2}
\end{eqnarray}
The configuration is depicted in Fig.~\ref{fig1}.

Since only a single degree of freedom is coupled to multiple
reservoirs, the system is maintained in a state described by a Boltzmann-like probability 
distribution  with  an effective temperature. This is in contrast to for example 
a harmonic chain coupled to two reservoirs, where a genuine 
non equilibrium situation is established, see e.g. \cite{Fogedby12}.
By inspection of the Langevin equation (\ref{lan1}) for $x$ we infer
the effective equilibrium temperature
\begin{eqnarray}
T=\frac{\sum_n\Gamma_nT_n}{\sum\Gamma_n},
\label{temp}
\end{eqnarray}
and the stationary distribution $P_0(x)\propto\exp(-U(x)/T)$. However,
the individual heat fluxes between the reservoirs via the particle constitute
a non equilibrium problem. From the equations of motion (\ref{lan1}) and
(\ref{lan2}) we also infer 
\begin{eqnarray}
\sum_nQ_n(t)=U(x)-U(x_0),
\label{tq}
\end{eqnarray}
expressing global energy conservation; here $x_0$ denotes the
initial configuration. We assume in the following that $Q(t_0)=0$, i.e.,
we start sampling the heat at $t=t_0$. In other words, whereas the
individual heat component $Q_n(t)$ is fluctuating, the sum $\sum_nQ_n(t)$
locks onto the nonfluctuating potential difference $U(x)-U(x_0)$.

The stochastic equations (\ref{lan1}) and (\ref{lan2}) define the
problem we wish to study. However, for the present purposes it is more convenient
to adhere to a Fokker-Planck description \cite{Risken89,vanKampen92}. 
Since the transfer of heat $\mathbf Q$ changes the state of the system and thus couples to 
the coordinate $x$, the distribution
for the coordinate and heat is characterised by the joint
distribution $P(\mathbf Q,xt)$ satisfying a Fokker-Planck equation
with Liouville operator $L(\mathbf Q,x)$ \cite{vanKampen92,Imparato07}, i.e.,
\begin{eqnarray}
&&\frac{dP(\mathbf Q,xt)}{dt}=L(\mathbf Q,x)P(\mathbf Q,xt),
\label{fp1}
\\
&&L(\mathbf Q,x)=\sum_n\Gamma_n\bigg[T_n\frac{d^2}{dx^2}+\frac{d}{dx}U'+(U'^2+T_nU'')\frac{d}{dQ_n}+
T_nU'^2\frac{d^2}{dQ_n^2}+2T_nU'\frac{d^2}{dxdQ_n}\bigg].~~~~~
\label{lop1}
\end{eqnarray}
Since $L(\mathbf Q,x)$is linear in $d/dQ_n$ it is convenient to introduce the
characteristic function $C_{\bm\lambda}(x,t)$ defined according to
\begin{eqnarray}
C_{\bm\lambda}(xt)=\int\prod_ndQ_n\exp(\bm\lambda\mathbf Q)P(\mathbf Q,xt),
\label{char1}
\end{eqnarray}
where $\bm\lambda\mathbf Q=\sum_n\lambda_nQ_n$. $C_{\bm\lambda}(x,t)$
then satisfy the Fokker-Planck equation with Liouville operator below
\begin{eqnarray}
&&\frac{dC_{\bm\lambda} (xt)}{dt}= L_{\bm\lambda}(x)C_{\bm\lambda}(xt),
\label{fp2}
\\
&&L_{\bm\lambda}(x)=\sum_n\Gamma_n\bigg[T_n\frac{d^2}{dx^2}+U'(x)(1-2\lambda_n T_n)\frac{d}{dx}+
(1-\lambda_n T_n)(U''(x)-\lambda_n U'^2(x))\bigg].~~~~
\label{lop2}
\end{eqnarray}
We note that setting $\lambda_n=0$ corresponds to integrating over $Q_n$ and
we have $C_{\bm\lambda=\mathbf 0}(xt)=P(xt)$, where $P(xt)$ is the distribution 
function for the coordinate $x$ in global equilibrium with the reservoirs at temperature 
$T$ given by (\ref{temp}). $P(xt)$ thus satisfies the Fokker-Planck equation
$dP(xt)/dt=L_\mathbf 0(x)P(xt)$, where $L_\mathbf 0(x)=\Gamma(Td^2/dx^2+d/dx U'(x))$,
with stationary distribution $P_0(x)\propto\exp(-U(x)/T)$.
\subsection{Symmetry}
\label{subsec:sym}
Incorporating the initial condition $x_0$ at time $t_0$ in the notation by setting 
$C_{\bm\lambda}(xt)\equiv C_{\bm\lambda}(xt,x_0t_0)$,
imposing the boundary condition $C_{\bm\lambda}(xt_0,x_0t_0)=\delta(x-x_0)$,
and introducing the matrix notation
\begin{eqnarray}
L_{\bm\lambda}(xx')=L_{\bm\lambda}(x)\delta(x-x'),
\label{mat}
\end{eqnarray}
the Fokker-Planck equation (\ref{fp2}) takes the form
\begin{eqnarray}
\frac{dC_{\bm\lambda} (xt,x_0t_0)}{dt}=\int dx' L_{\bm\lambda}(xx')C_{\bm\lambda}(x't,x_0t_0).
\label{fp3aux}
\end{eqnarray}
In analogy with corresponding manipulations for the evolution operator in
quantum mechanics \cite{Zinn-Justin89}, we obtain the formal solution  
\begin{eqnarray}
C_{\bm\lambda}(xt,x_0t_0)=\Big[\exp(L_{\bm\lambda}(t-t_0))\Big](x,x_0);
\label{char2}
\end{eqnarray}
here the the n-th term in the expansion of the exponential is defined according to
$(1/n!)(t-t_0)^n\int dx_1\cdots dx_{n-1}L_{\bm\lambda}(x,x_1)\cdots L_{\bm\lambda}(x_{n-1},x_0)$.

Choosing a particular reservoir at temperature $T_k=1/\beta_k$  inspection of 
 the Liouville operator (\ref{lop2}) in the form (\ref{mat}) yields the symmetry
\begin{eqnarray}
L_{\{\lambda_n\}}(x,x')=\exp(-\beta_k(U(x)-U(x')))L_{\{\beta_{nk}-\lambda_n\}}(x',x),
\label{lopsym}
\end{eqnarray}
for details see Appendix B. From the solution (\ref{char2}) we then readily infer
the corresponding symmetry for $C_{\{\lambda_n\}}$, i.e.,
\begin{eqnarray}
C_{\{\lambda_n\}}(xt,x_0t_0)=
\exp(-\beta_k(U(x)-U(x_0)))C_{\{\beta_{nk}-\lambda_n\}}(x_0t;xt_0).
\label{csym}
\end{eqnarray}
which holds for any choice of $k$.
From the inverse Laplace transform  (\ref{char1}) \cite{Lebedev72},
\begin{eqnarray}
P(\mathbf Q,xt,x_0t_0)=
\int_{-i\infty}^{i\infty}\prod_n\frac{d\lambda_n}{2\pi i}\exp(-\bm\lambda\mathbf Q)C_{\bm\lambda}(xt,x_0t_0),
\label{p1}
\end{eqnarray}
we finally obtain (\ref{psym}), i.e.,
\begin{eqnarray}
P(\mathbf Q,xt,x_0t_0)=\exp(-\beta_k(U(x)-U(x_0)))\exp\Big(-\sum_n\beta_{nk}Q_n\Big)P(-\mathbf Q,x_0t,xt_0),
\label{psymaux}
\end{eqnarray}
that again holds for all $k$.
In the special case $\beta_k=0$ we obtain in particular the simpler symmetries
\begin{eqnarray}
&&C_{\{\lambda_n\}}(xt,x_0t_0)=C_{\{\beta_n-\lambda_n\}}(x_0t,xt_0),
\label{csym1}
\\
&&P(\mathbf Q,xt,x_0t_0)=\exp\Big(-\sum_n\beta_{n}Q_n\Big)P(-\mathbf Q,x_0t,xt_0).
\label{psym1}
\end{eqnarray}
This concludes the demonstration of the symmetries for the characteristic function
and the joint heat distribution. In the next section we consider the implications
for the cumulant generating function.
\section{\label{cum}Cumulant generating function}
\subsection{General}
At long time the characteristic function $C_{\{\lambda_n\}}(xt,x_0t_0)$
is dominated by the largest eigenvalue $\mu_0(\bm\lambda)$ of the 
Liouville operator $L_{\bm\lambda}$. This follows heuristically from
the solution (\ref{char2}) and can be made more precise by considering
the spectral representation of $L_{\bm\lambda}(x,x_0)$; this discussion
is deferred to Appendix A. We have at long times compared to $t_0$
\begin{eqnarray}
C_{\bm\lambda}(xt,x_0t_0)=A_0(xx_0,\bm\lambda)\exp(\mu_0(\bm\lambda)t),
\label{char3}
\end{eqnarray}
where $\mu_0(\bm\lambda)$ is the cumulant generating function.
The prefactor $A_0(xx_0,\bm\lambda)$ is expressed in terms of the eigenfunctions associated with the
eigenvalue $\mu_0(\bm\lambda)$, for details see Appendix A.
From the definition (\ref{char1}) we also have, denoting the constrained
average by the subscript $xx_0$,
\begin{eqnarray}
\langle\exp(\bm\lambda\mathbf Q(t))\rangle_{xx_0}=A(xx_0,\bm\lambda)\exp(\mu_0(\bm\lambda)t),
\label{char4}
\end{eqnarray}
and it follows that $\mu_0$ yields the mean heat and higher cumulants,
i.e. $\langle Q_n(t)\rangle_{xx_0}/t=A(xx_0,\mathbf 0)(d\mu/d\lambda_n)_{\bm\lambda=0}$,
$(\langle Q(t)^2\rangle_{xx_0}-(\langle Q(t)\rangle_{xx_0})^2)/t=
A(xx_0,\mathbf 0)(d^2\mu/d\lambda_n^2)_{\bm\lambda=0}$, etc.

Applying the symmetry (\ref{csym1}) to (\ref{char3}) we readily obtain the usual asymptotic
fluctuation theorem \cite{Gallavotti95,Lebowitz99} generalised to multiple reservoirs
\begin{eqnarray}
\mu_0(\{\lambda_n\})=\mu_0(\{\beta_n-\lambda_n\}).
\label{cgf1}
\end{eqnarray}
Likewise,  we infer, applying the full symmetry (\ref{csym}) to (\ref{char3}), the generalised fluctuation theorem
$\mu_0(\{\lambda_n\})=\mu_0(\{\beta_{nk}-\lambda_n\}), \beta_{nk}=\beta_n-\beta_k$,
for fixed $k$. Here we note that since $Q_n(t)\sim t$ the  exponential $\exp(-\beta_k(U(x)-U(x_0))$ in
(\ref{csym}) is subdominant in time compared with $\exp(\sum_n(\beta_{nk}-\lambda_n)Q_n(t))$ at long times.
Moreover, since this symmetry holds for any choice of $k$ we conclude
that $\mu_0$ only depends on the difference variable
\begin{eqnarray}
\nu_{nm}=\lambda_n-\lambda_m,
\label{diff}
\end{eqnarray}
and we have  
\begin{eqnarray}
\mu_0(\{\nu_{nm}\})=\mu_0(\{\beta_{nm}-\nu_{nm}\}).
\label{cgf2}
\end{eqnarray}
This relation generalises the usual asymptotic fluctuation theorem to the case of
multiple reservoirs.

From (\ref{p1}) the heat distribution at long times is given by
\begin{eqnarray}
P(\mathbf Q,xt,x_0t_0)=
\int_{-i\infty}^{i\infty}\prod_n\frac{d\lambda_n}{2\pi i}A(xx_0,\bm\lambda)\exp(-\bm\lambda\mathbf Q+\mu_0(\bm\lambda)t).
\label{p2}
\end{eqnarray}
Since the integral at long times is controlled by the exponential and $\mu_0$ depends 
on the difference $\lambda_n-\lambda_m$, it follows that the integral is invariant under the
shift $\lambda_n\rightarrow\lambda_n+a$, where $a$ is arbitrary. Consequently,
$P(\mathbf Q,xt,x_0t_0)=\exp(-a\sum_n Q_n))P(\mathbf Q,xt,x_0t_0)$ and we obtain
the constraint 
\begin{eqnarray}
\sum_nQ_n=0;
\label{con}
\end{eqnarray}
note that this constraint on the heat variables $\mathbf Q$ in the heat distribution 
$P(\mathbf Q,xt,x_0t_0)$ at long times is not in conflict with (\ref{tq}), i.e.
$\sum_nQ_n(t)=U(x)-U(x_0)$ which expresses energy conservation for the
fluctuating heat. At long times the dependence of $P$ on $Q_n$ is confined to the sub manifold
determined by (\ref{con}).

The condition (\ref{con}) also follows from the symmetry (\ref{psym})
noting that $\exp(-\beta_k(U(x)-U(x_0)))$ is subdominant in time and that $k$ can be
chosen arbitrarily. We have in abbreviated notation
\begin{eqnarray}
P(Q_1,\cdots Q_N,t)=\delta(Q_1+\cdots+Q_N)P(Q_1,\cdots Q_{N-1},t).
\label{p3}
\end{eqnarray}
In other words, for $N$ reservoirs there are only $N-1$ independent heat transfers
as a result of the constraint in (\ref{con}). For two reservoirs we have for example
$P(Q_1,Q_2,t)=\delta(Q_1+Q_2)P_1(Q_1,t)$, where $P_1(Q,t)$ is the heat 
distribution for reservoir 1 and we infer that 
$P_2(Q,t)=P_1(-Q,t)$, i.e., the distribution for the outgoing heat  from reservoir 1 is
identical to the distribution for incoming heat to reservoir 2, expressing conservation of
energy.

By the usual steepest descent argument \cite{Touchette09,Lebowitz99,Fogedby12} the prefactor
locks onto $A(xx_0,\bm\lambda^\ast)$, where $\lambda_n^\ast$ is a solution of
$Q_n/t=(\partial\mu/\partial\lambda_n)_{\lambda_n=\lambda_n^\ast}$ and we have
the scaling form
\begin{eqnarray}
&&P(\mathbf Q,xt,x_0t_0)
\propto A(xx_0,\bm\lambda^\ast)\exp(-F(\mathbf Q/t)t),
\label{pst}
\\
&&F(\mathbf Q/t)=-\mu_0(\bm\lambda^\ast)+\bm\lambda^\ast \mathbf Q/t,
\label{ldf}
\end{eqnarray}
where $F(\mathbf q)$, $\mathbf q=\mathbf Q/t$, is the large deviation function in the case
of multiple reservoirs. The symmetry (\ref{psym}) applied to $F(\mathbf q)$ 
then yields the fluctuation theorem for the large deviation function
\begin{eqnarray}
F(\mathbf q)=\sum_n\beta_{nk}q_n+F(-\mathbf q)~~\text{for arbitrary}~~k. 
\label{ft2}
\end{eqnarray}
%
\subsection{Branch points and tails}
It follows from general principles \cite{Touchette09,Hollander00} that the cumulant generating 
function $\mu_0(\bm\lambda)$ is bounded and possesses a branch cut structure.
For large $\mathbf Q$ the Laplace transform in (\ref{p2}), when closing the contour
in its evaluation, samples the edges of the branch cuts closest to the origin determining
the tails of the heat distribution. We note that in principle the singular structure of the prefactor
$A(xx_0,\bm\lambda)$ can also influence the tails. This issue has been discussed in detail for the
case of a particle driven by two heat reservoirs \cite{Visco06,Farago02}. In the present 
context we only analyse the contribution arising from the singular structure of $\mu_0(\bm\lambda)$.
In general  $\mu_0(\bm\lambda)$ is a downward convex function passing through
the origin $\mu_0(\mathbf 0)=0$ due to normalzation. From the symmetry (\ref{cgf2}) it
then also follows that $\mu_0(\{\beta_{nm}\})=0$. Moreover, if $\mu_0$ has a branch point
at $\nu_{nm}^{\text{BP}}$ there will also be a branch point present at $\beta_{nm}-\nu_{nm}^{\text{BP}}$.

In order to be more specific we return to the model in Sec. \ref{sys} and note that the 
similarity transformation $\exp(g)$ \cite{Imparato07}, where
\begin{eqnarray}
g(x,\bm\lambda)=U(x)\frac{\sum_n\Gamma_n T_n(\lambda_n-\beta_n/2) }{\Gamma T},
\label{g}
\end{eqnarray}
maps the Liouville operator $L_{\bm\lambda}$ in (\ref{lop2}) to a Hermitian Schr\"odinger
form, i.e., $L^S_{\bm\lambda}=\exp(-g)L_{\bm\lambda}\exp(g)$, where
\begin{eqnarray}
&&L^S_{\bm\lambda}(x)=
\Gamma\left[T\frac{d^2}{dx^2}+\frac{1}{2}U''-\frac{1}{4T}\left(\frac{\omega}{\Gamma}\right)^2U'^2\right].
\label{lop3}
\end{eqnarray} 
Here
\begin{eqnarray}
&&\omega^2=\sum_{nm}\Gamma_n\Gamma_m[1-F_{nm}],
\label{om}
\\
&&F_{nm}=2\nu_{nm}(T_n-T_m+T_nT_m\nu_{nm}).
\label{f}
\end{eqnarray} 
Consulting Appendix C it follows that  $L_{\bm\lambda}$ and $L^S_{\bm\lambda}$
have identical spectra, in particular the same largest eigenvalue $\mu_0$.
From the structure of $L^S_{\bm\lambda}$ it, moreover, follows that $\mu_0$
depends on the form of $U$ and parametrically on the temperature dependent 
parameter $\omega$. We also note that the function $F_{nm}$ is invariant under
the transformation $\nu_{nm}\rightarrow\beta_{nm}-\nu_{nm}$, in accordance with
(\ref{cgf2})

In order to obtain an explicit expression for $\mu_0$ we consider in the following
a harmonic oscillator potential $U(x)=x^2/2$. Since $U''(x)=1$ and $U'(x)=x$ the operator
(\ref{lop3}) corresponds to a quantum oscillator with mass $1/2T$ moving in the 
potential $(1/4T)(\omega/\Gamma)^2x^2-1/2$. Referring to Appendix D the discrete
spectrum is given by $\mu_n(\{\nu_{nm}\})=(1/2)(\Gamma-\omega)-\omega n$, 
$n=0,1,\cdots$, and we obtain in particular the cumulant generating function 
$\mu_0(\omega)=(1/2)(\Gamma-\omega)$. In more detail inserting (\ref{om})
\begin{eqnarray}
\mu_0(\{\nu_{nm}\})=
\frac{1}{2}\Bigg(\Gamma-\Big(\sum_{nm}\Gamma_n\Gamma_m[1-F_{nm}]\Big)^{1/2}\Bigg).
\label{cgf3}
\end{eqnarray} 
%
\subsubsection{One reservoir}
The situation is simple in the case of a single reservoir. We have
$\lambda_n=\lambda$, i.e., $\nu_{nm}=0$ and from (\ref{om}) $F_{nm}=0$.
$\omega$ locks onto $\Gamma^2$ and we have $\mu_0=0$ for all $\lambda$,
i.e., a vanishing cumulant generating function. This is consistent with the observation
that a single degree of freedom coupled to a single heat reservoir is maintained 
in equilibrium at temperature $T$ with a bounded heat distribution 
\cite{Imparato07,Fogedby11a}. 
\subsubsection{Two reservoirs}
The case of two reservoirs with dampings $\Gamma_1$ and $\Gamma_2$
and temperatures $T_1$ and $T_2$ was  considered by Derrida et al.
\cite{Derrida05}, see also \cite{Fogedby11a,Visco06}. Here we have 
$\nu=\lambda_1-\lambda_2$, $F_{12}=2\nu(T_1-T_2+T_1T_2\nu)$,
$\omega^2=\Gamma^2-2\Gamma_1\Gamma_2F_{12}$, and the 
cumulant generating function is given by the explicit expression
\begin{eqnarray}
\mu_0(\nu)=\frac{1}{2}\bigg(\Gamma-\bigg(\Gamma^2+4\Gamma_1T_1\Gamma_2T_2\nu(\beta_{12}-\nu)\bigg)^{1/2}\bigg).
\label{cgf4}
\end{eqnarray} 
The cumulant generating function is a downward convex function
passing through $\mu_0(0)=0$ and $\mu_0(\beta_{12})=0$. It, moreover, 
obeys the symmetry $\mu_0(\nu)=\mu_0(\beta_{12}-\nu)$. 
Closing off reservoir 2 by setting $\Gamma_2=0$ we have $\mu=0$
since a single particle driven by one reservoir is in equilibrium.
The branch points are given by \cite{Fogedby11a}
\begin{eqnarray}
\nu_\pm=\frac{1}{2}\bigg(\beta_{12}\pm\bigg(\beta_{12}^2+\Gamma^2/\Gamma_1T_1\Gamma_2T_2\bigg)^{1/2}\bigg).
\label{bp1}
\end{eqnarray} 
In the equal temperature case for $T_1=T_2=T$, i.e., $\beta_{12}=0$ we have in particular
the branch points $\nu_\pm=\pm (1/2T)((\Gamma_1+\Gamma_2)/\sqrt{\Gamma_1\Gamma_2})$.
For $\Gamma_1=\Gamma_2$ the branch points are independent of the damping
and located at 
\begin{eqnarray}
\nu_\pm=\pm\frac{1}{T}.
\label{bp2}
\end{eqnarray} 
%
\subsubsection{Multiple reservoirs}
For multiple reservoirs the analysis is based on the general 
expression in (\ref{cgf3}) originating from the harmonic potential case. 
We, moreover, for simplicity consider the case of $N$ reservoirs
with identical damping constants, i.e., $\Gamma_n=\tilde\Gamma$ and 
thus $\Gamma=N\tilde\Gamma$. We find
\begin{eqnarray}
\mu_0(\{\nu_{nm}\})=\frac{\tilde\Gamma}{2}\Big(N-\Big(\sum_{nm}[1-F_{nm}]\Big)^{1/2}\Big),
\label{cgf8}
\end{eqnarray} 
and the branch point condition reads  $\sum_{nm}[1-F_{nm}]=0$. This is still
a complex expression to analyse; however, focussing on reservoir 1
by setting $\lambda_1=\lambda$ and $\lambda_n=0$, 
for $n\neq 1$ we have $\nu_{1n}=\lambda$ for $n=2,3,\cdots, N$ and we  
obtain the branch points
\begin{eqnarray}
\nu_\pm=\pm\frac{N}{2\sqrt{N-1}}\frac{1}{T}.
\label{bp3}
\end{eqnarray} 
For $N=2$ we recover (\ref{bp2}); for $N=3$ we have 
$\nu_\pm=\pm(1/T)3/2\sqrt{2}$. As the number of reservoirs
increase the branch points recede to infinity as $\sim \sqrt{N}/2T$.
\subsection{Heat distribution}
For small $\lambda_n$ and assuming $\beta_{nm}$ small we can replace
$\mu_0$ by a parabolic approximation consistent with the symmetry 
(\ref{cgf2}) and the boundary condition $\mu_0(\mathbf 0)=0$
i.e.,
\begin{eqnarray}
\mu_0(\{\nu_{nm}\})=\sum_{nm}a_{nm}\nu_{nm}(\beta_{nm}-\nu_{nm}).
\label{par}
\end{eqnarray} 
Here $a_{nm}$ is symmetric and related to the mean heat flux according to
$\langle q_p\rangle=2\sum_n a_{pn}\beta_{pn}$. From the steepest descent
calculation we then obtain after some algebra the following expression for 
the large deviation function
\begin{eqnarray}
F(\{\lambda_n\})=\sum_{nm}a_{nm}\nu_{nm}^\ast(\beta_{nm}-\nu_{nm}^\ast)-\sum_m\lambda_m^\ast q_m, 
\label{fpar}
\end{eqnarray} 
where 
\begin{eqnarray}
&&\lambda_n^\ast=\frac{1}{2}\sum_pK_{np}\Big(\sum_m a_{pm}\beta_{pm}-\frac{q_p}{2}\Big),
\label{s1}
\\
&&K_{np}=\Big(\delta_{np}\sum_q a_{pq}-a_{np}\Big)^{-1}.
\label{s2}
\end{eqnarray} 
Note that $F$ is quadratic in $q_n$ yielding a Gaussian
heat distribution for small $q_n$;
for the corresponding analysis in the case of two reservoirs,
see \cite{Fogedby12}.

Regarding the tails in the heat distribution the transform in (\ref{p2})
samples for large $q_n$ the branch points in $\mu_0$. Focussing on the 
heat distribution for reservoir 1 in the presence of the other $N-1$ reservoirs  we
thus obtain
\begin{eqnarray}
&&P(q_1)\sim\exp(-\nu_+q_1t)~~\text{for large positive $q_1$},
\\
&&P(q_1)\sim\exp(-|\nu_-||q_1|t)~~\text{for large negative $q_1$},
\end{eqnarray} 
where $\nu_\pm$ for $N$ reservoirs is given by (\ref{bp3}).
\section{\label{fluc}Fluctuation theorems}
\subsection{Exact fluctuation theorem}
Here we discuss further consequences of the symmetry (\ref{psym})
valid at all times. Multiplying both sides with $\exp(-\beta_kU(x_0))$, integrating
over $x$ and $x_0$, and exchanging $x$ and $x_0$ in the integral
on the left hand side, we obtain the identity
\begin{eqnarray}
P^{(k)}(\mathbf Q,t,t_0)=\exp\Big(-\sum_n\beta_{nk}Q_n\Big)P^{(k)}(-\mathbf Q,t,t_0),
\label{ft}
\end{eqnarray}
where we have defined the average
\begin{eqnarray}
P^{(k)}(\mathbf Q,t,t_0)=\int dxdx_0P(\mathbf Q,x_0t,xt_0)
\exp(-\beta_kU(x_0)).
\label{paver}
\end{eqnarray}
Here $P^{(k)}(\mathbf Q,t,t_0)$ is the joint heat distribution associated
with a non equilibrium transition from $x_0$ at time $t_0$ to $x$ at time
$t$ with the system in equilibrium with the $k$-th reservoir at the initial time $t_0$.
The expression (\ref{ft}) is a fluctuation theorem valid at all times and is a particular case of a more general expression  
first derived by Cuetara et al. \cite{Cuetara14} for a system in contact with several energy and particle reservoirs, on the basis of entropic 
considerations at the level of single trajectories.

In the case of two reservoirs setting $k=2$ we obtain in particular
for the joint heat distribution
$P^{(2)}(Q_1,Q_2,t,t_0)=\exp(-(\beta_1-\beta_2)Q_1)P^{(2)}(-Q_1,-Q_2,t,t_0)$
or integrating over  $Q_2$ the fluctuation theorem
\begin{eqnarray}
P^{(2)}(Q_1,t,t_0)=\exp(-(\beta_1-\beta_2)Q_1)P^{(2)}(-Q_1,t,t_0).
\label{ft211}
\end{eqnarray}
The interpretation is straightforward, see \cite{Cuetara14}.
Preparing the system in equilibrium at time $t_0$ with the reservoir at temperature
$T_2$ and subsequently monitoring the heat transferred from the
reservoir at temperature $T_1$, the heat distribution obeys the fluctuation theorem in
(\ref{ft211}) valid at all times.
 
In the case of multiple reservoirs we prepare the system in equilibrium 
with the k-th reservoir at temperature $T_k$ at time $t_0$ and the
fluctuation theorem (\ref{ft}) applies to the joint heat distribution
for the other reservoirs. Note that whereas the asymptotic
fluctuation theorem (\ref{cgf2}) for the cumulant generating function is generic
in the sense that it is independent of the potential $U(x)$, the fluctuation theorem 
in the present context requires that the system
is in equilibrium with distribution $\exp(-\beta_k U(x_0))$ with one of the reservoirs at the initial time.
\subsection{Trajectory interpretation}
\subsubsection{Time independent potential - no work protocol}
In order to establish contact with the trajectory point of view pursued by Seifert
 \cite{Seifert05a}, based on the stochastic thermodynamics scheme by
 Sekimoto \cite{Sekimoto98}, and the role of
entropy changes in the course of the non equilibrium time evolution,
we express the definition of the characteristic function
in the form
\begin{eqnarray}
C_{\bm\lambda}(xt,x_0t_0)=\langle\exp\Big(\sum_n\lambda_n Q_n(t)\Big)\delta(x-f(t,x_0t_0))\rangle.
\label{defc}
\end{eqnarray}
Here $f(t,x_0t_0)$ is a solution of the Langevin equation (\ref{lan1}) for a specific noise
realisation $\xi(t)$ defining a forward trajectory in configuration space from the initial configuration
$x_0$ at time $t_0$ to the final configuration  $x$ at time $t$. Likewise, solving
$x=f(t,x_0t_0)$ for $x_0$, i.e., $x_0=\tilde f(t,xt_0)$, we identity a backward trajectory 
from $x$ at time $t_0$ to $x_0$ at time $t$. The symmetry (\ref{csym1}) then reads
\begin{eqnarray}
\langle\exp\Big(\sum_n\lambda_n Q_n(t)\Big)\delta(x-f(t,x_0t_0))\rangle=
\langle\exp\Big(\sum_n(\beta_n-\lambda_n) Q_n(t)\Big)\delta(x_0-\tilde f(t,xt_0))\rangle.
\label{symc}
\end{eqnarray}
For a particular noise realisation defining a trajectory in configuration space we obtain 
\begin{eqnarray}
\exp\Big(\sum_n\lambda_n Q^{\text{F}}_n(t)\Big)\delta(x-f(t,x_0t_0))=
\exp\Big(\sum_n(\beta_n-\lambda_n) Q^{\text{B}}_n(t)\Big)\delta(x_0-\tilde f(t,xt_0)).
\label{symtraj}
\end{eqnarray}
Note that $Q^{\text{F}}_n(t)$ is a solution of the
Langevin equation (\ref{lan2}). Since $\delta(x-f(t,x_0t_0))$ defines a forward trajectory
the heat transfer $Q^{\text{F}}_n(t)$ is associated with this trajectory; likewise, $Q^{\text{B}}_n(t)$
is associated with a backward trajectory. For the special choice $\lambda_n=\beta_n$
we obtain in particular
\begin{eqnarray}
\exp\Big(\sum_n\beta_n Q^{\text{F}}_n(t)\Big)\delta(x-f(t,x_0t_0))=\delta(x_0-\tilde f(t,xt_0)).
\label{symtraj1}
\end{eqnarray}
This relationship is completely equivalent to the analysis in \cite{Seifert05a} based on
the path integral formulation of the solution to the Fokker-Planck for the distribution
$P(xt,x_0t_0)$.

Setting  $\lambda_n=\beta_n$ in (\ref{symc}) and integrating over the initial
position $x_0$ we obtain 
\begin{eqnarray}
\int dx_0\langle\exp\Big(\sum_n\beta_n Q_n(t)\Big)\delta(x-f(t,x_0t_0))\rangle=1,
\label{symc2}
\end{eqnarray}
where we have used the normalisation condition $\int dx_0\langle\delta(x_0-\tilde f(t,xt_0))\rangle=1$
or, equivalently, $\int dx_0P(x_0t,xt_0)=1$, i.e., the conservation of probability.

Introducing the total increase of the heat bath entropy
\begin{eqnarray}
S_{\text{bath}}(t)=-\sum_n\beta_nQ_n(t),
\label{entres}
\end{eqnarray}
we can also express (\ref{symc2}) in the form of a fluctuation theorem
for the bath entropy
\begin{eqnarray}
\int dx_0\langle\exp(-S_{\text{bath}}(t))\rangle_{xt,x_0t_0}=1.
\label{symc3}
\end{eqnarray}
This fluctuation theorem states that integrated over the
initial configuration $x_0$ with weight one, the constrained 
average of $\exp(-S_{\text{bath}}(t))$ along  forward trajectories equals unity. 
The fluctuation theorem relates to the total reservoir entropy production 
in the course of a non equilibrium transition. The fluctuation theorem
is a consequence of a basic symmetry of the joint heat-position distribution
combined with the normalisation of the distribution for the position.
We also note that the joint distribution incorporates on the Fokker-Planck
level the stochastic thermodynamics scheme developed within a Langevin
formulation \cite{Sekimoto98,Seifert05a}.
\subsubsection{Time dependent potential - with work protocol}
In the case of a work protocol characterised by a time dependent potential
$U(xt)$ modelling work performed on the system, the symmetries in
(\ref{csym}) and (\ref{csym1}) are not available since the Liouville
operator acquires an explicit time dependence.  The extension of the
scheme to the time dependent case is summarised in Appendix E.
Whereas the fundamental symmetry still applies to the Liouville operator,
i.e.,
\begin{eqnarray}
L_{\{\lambda_n\}}(xx',t)=\exp(-\beta_k(U(xt)-U(x't)))L_{\{\beta_{nk}-\lambda_n\}}(x'x,t),
\label{symap}
\end{eqnarray}
the characteristic function is given by the time ordered product \cite{Zinn-Justin89},
\begin{eqnarray}
C_{\bm\lambda}(xt,x't')=
T\Big[\exp\Big(\int_{t'}^t dt'' \hat L_{\bm\lambda}(t'')\Big)\Big](xx').
\label{time2ap}
\end{eqnarray}
Considering for example the second order term in the expansion we obtain
applying the symmetry
\begin{eqnarray}
&&C_{\{\lambda_n\}}^{(2)}(xt,x't')= 
\int_{t'}^{t}dt_1\int_{t'}^{t_1}dt_2\int dyL_{\{\lambda_n\}}(xy,t_1)L_{\{\lambda_n\}}(yx',t_2)=
\nonumber
\\
&&\int_{t'}^{t}dt_1\int_{t'}^{t_1}dt_2\int dyL_{\{\beta_{nk}-\lambda_n\}}yx,t_1)L_{\{\beta_{nk}-\lambda_n\}}(x'y,t_2)
e^{-\beta_k(U(xt_1)-U(yt_1)+U(yt_2)-U(x't_2)},~~~~~
\label{exp2}
\end{eqnarray}
and we note that the time dependent protocol as exemplified by the time dependence of $U$
becomes entirely entangled in the time integrations. On the contrary, in the time independent 
case the factors $\exp(-\beta_kU(x))$ in the expansion of the exponential in (\ref{time2ap}) 
cancel and we recover the symmetry in (\ref{csym}).

We also note in passing
that in the case of a moving  harmonic potential of the form $U(xt)\propto (x-vt)^2$, as discussed
in \cite{Imparato07}, the time ordered expression reduces to integration
over Gaussians (yielding error functions) and can possible be reduced. However, we abstain from 
such an analysis in the present context; we refer to \cite{Imparato07} for a
steepest descent analysis.

With respect to the bath-fluctuation theorem (\ref{symc3}) we note that setting $\beta_k=0$
we have in operator form, see Appendix E,
\begin{eqnarray}
C_{\{\lambda_n\}}(xt,x't')=
T\Big[\exp\Big(\int_{t'}^t dt'' \hat L_{\{\beta_n-\lambda_n\}}^\ast(t'')\Big)\Big](xx'),
\label{time3ap}
\end{eqnarray}
and in particular setting $\beta_n=\lambda_n$
\begin{eqnarray}
C_{\{\beta_n\}}(xt,x't')=
T\Big[\exp\Big(\int_{t'}^t dt'' \hat L_{\{0\}}^\ast(t'')\Big)\Big](xx').
\label{time4ap}
\end{eqnarray}
It follows from the form of the Liouville operator in (\ref{lopt2}) that
$\int dx L_{\{0\}}(xx',t)=0$, expressing conservation of probability
in the Fokker-Planck equation for $P(xt,x't')$, $dP(xt,x't')/dt=
\int dx''L_{\{0\}}(xx'',t)P(x''t,x't')$. Correspondingly, for the transposed
Liouvillian $L^\ast$ we have $\int dx' L_{\{0\}}^\ast(xx',t)=0$ and it follows
from (\ref{time4ap}) that also in the time dependent case do we obtain
(\ref{symc2}), yielding the fluctuation theorem (\ref{symc3}).

The above analysis might appear cumbersome and possibly superfluous. 
General physical arguments imply that the fluctuation
theorem for the bath entropy only monitors the transfer of heat and thus does not
depend on the applied work protocol.
\subsection{Integral fluctuation theorem}
Seifert has proposed an integral fluctuation theorem valid at all times
based on entropy production and consumption on the trajectory level 
\cite{Seifert05a}. Within the present context the integral fluctuation 
theorem follows readily from (\ref{symc3}) and is therefore a consequence 
of the basic symmetry of the Liouville operator.

The bath fluctuation theorem was formulated as an integral condition
for the constrained average of the fluctuating heat or entropy $-S_{\text{bath}}(t)$
for an initial configuration at $x_0$ at time $t_0$ and a final configuration $x$
at time $t$. In order to make contact with the integral fluctuation theorem 
one must consider fluctuating initial and final configuration characterised 
by the normalised distributions $P_t(x)$ and $P_0(x_0)$. 

Based on the path integral formulation of the distribution $P(xt,x't')$ it was shown in 
\cite{Seifert05a} that one can define entropy production and consumption
for an individual trajectory from $x_0$ at time $t_0$ to $x$ at time $t$
based on the Gibbs expression $S=-\sum_n p_n\ln p_n$. By inspection of the
entropy associated with the forward and backward trajectories one easily
extracts the entropy associated with the contact with the heat baths, 
$S_{\text{bath}}$. The additional entropy associated with the non equilibrium transition itself
on the trajectory level is then given in terms of the initial and final
distribution as $S_{\text{sys}}=\ln P_0(x_0)-\ln P_t(x)$.

Averaging over the initial distribution $P_0(x_0)$ 
we infer from (\ref{symc3})
\begin{eqnarray}
\int dx_0\langle\exp(-S_{\text{bath}}(t)-\ln P_0(x_0))\rangle_{xt,x_0t_0}P_0(x_0)=1,
\label{symc4}
\end{eqnarray}
where the distribution $P_0(x_0)$ is balanced by the  entropy
contribution $\ln P_0(x_0)$. Finally, integrating over the final state
$x$ with the normalised weight $P_t(x)$, i.e., $\int dx P_t(x)=1$,
we obtain the integral fluctuation theorem proposed by Seifert \cite{Seifert05a}
\begin{eqnarray}
\int dx \int dx_0\langle\exp(-S_{\text{total}}(t))\rangle_{xt,x_0t_0}P_0(x_0)=1,
\label{symc5}
\end{eqnarray}
or
\begin{eqnarray}
\langle\exp(-S_{\text{total}}(t))\rangle=1,
\label{symc6}
\end{eqnarray}
where the total entropy is given by
\begin{eqnarray}
&&S_{\text{total}}= S_{\text{bath}}(t)+S_{\text{sys}},
\label{stot}
\\
&&S_{\text{sys}}=\ln P_0(x_0)-\ln P_t(x).
\label{ssys}
\end{eqnarray}
%
\section{\label{ham}Reservoir-driven Hamiltonian system}
The last issue to be considered in this section is the case of a system with 
many degrees of freedom. We consider an over damped system coupled 
to multiple reservoirs described by the Hamiltonian $H(\mathbf x\mathbf p)$. 
For the equations of motion for $\mathbf x$, $\mathbf p$ and $\mathbf Q$
we obtain
\begin{eqnarray}
&&\frac{dx_n}{dt}=p_n,
\label{eq1}
\\
&&\frac{dp_n}{dt}=-\frac{dH}{dx_n}-\Gamma_n p_n+\xi_n,
\label{eq2}
\\
&&\langle \xi_n \xi_m\rangle(t)=2\Gamma_n T_n\delta_{nm}\delta(t),
\label{n2}
\\
&&\frac{dQ_n}{dt}=p_n(-\Gamma_np_n+\xi_n),
\label{q2}
\end{eqnarray} 
and the conservation law
\begin{eqnarray}
\sum_nQ_n(t)=H(\mathbf x\mathbf p)-H(\mathbf x_0\mathbf p_0),
\label{cl}
\end{eqnarray} 
where $(\mathbf x_0,\mathbf p_0)$ is the initial phase space point at time $t_0$
and $(\mathbf x,\mathbf p)$ the phase space point at time $t$.

The Fokker-Planck equation for the joint distribution $P(\mathbf Q,\mathbf x\mathbf pt)$
has the form
\begin{eqnarray}
\frac{dP(\mathbf Q,\mathbf x\mathbf pt)}{dt}&=&(L_0(\mathbf x\mathbf p)+
L_{\mathbf Q}(\mathbf p))P(\mathbf Q,\mathbf x\mathbf pt),
\label{fp3}
\\
L_0(\mathbf x\mathbf p)&=&\sum_n\bigg[\left(\frac{dH}{dx_n}\frac{d}{dp_n}-\frac{dH}{dp_n}\frac{d}{dx_n}\right)
+\Gamma_n\left(T_n\frac{d^2}{dp_n^2}+\frac{d}{dp_n}p_n\right)\bigg],
\label{lpo31}
\\
L_{\mathbf Q}(\mathbf p)&=&\sum_n\Gamma_n\bigg[T_np_n^2\frac{d^2}{dQ_n^2}+
2T_np_n\frac{d^2}{dQ_ndp_n}
+(p_n^2+T_n)\frac{d}{dQ_n}\bigg];
\label{lop32}
\end{eqnarray} 
we note the appearance of  a Poisson bracket accounting for the deterministic dynamics.
Likewise, the characteristic function 
\begin{eqnarray}
C_{\bm\lambda}(\mathbf x\mathbf pt)=\int\prod_ndQ_n\exp(\bm\lambda\mathbf Q)
P(\mathbf Q,\mathbf x\mathbf pt),
\label{charham}
\end{eqnarray} 
is governed by the corresponding Fokker-Planck equation
\begin{eqnarray}
\frac{dC_{\bm\lambda}(\mathbf x\mathbf pt)}{dt}&=&L_{\bm\lambda}(\mathbf x\mathbf p)
C_{\bm\lambda}(\mathbf x\mathbf p,t),
\label{fp4}
\\
L_{\bm\lambda}(\mathbf x\mathbf p)&=&\sum_n\bigg(\frac{dH}{dx_n}\frac{d}{dp_n}-\frac{dH}{dp_n}\frac{d}{dx_n}\bigg)
\nonumber
\\
&&+\sum_n\Gamma_n\bigg[T_n\frac{d^2}{dp_n^2}+(1-2\lambda_nT_n)p_n\frac{d}{dp_n}
+(1-\lambda_nT_n)(1-\lambda_np_n^2)\bigg].
\label{lop4}
\end{eqnarray} 
Setting $L_{\bm\lambda}(\mathbf x\mathbf p,\mathbf x'\mathbf p')=
L_{\bm\lambda}(\mathbf x\mathbf p)\delta(\mathbf x-\mathbf x')
\delta(\mathbf p-\mathbf p')$ and including a time reversal operation
applied to the Poisson bracket in accordance with microscopic
reversibility, we note that the momentum as a result will change sign,
i.e., $\mathbf p\rightarrow-\mathbf p$. By inspection we find  the fundamental  
symmetry analogous to (\ref{lopsym}),
\begin{eqnarray}
L_{\{\lambda_n\}}(\mathbf x\mathbf p;\mathbf x'\mathbf p')=
\exp(-\beta_k(H(\mathbf x\mathbf p)-H(\mathbf x'\mathbf p')))
L_{\{\beta_{nk}-\lambda_n\}}(\mathbf x'(-\mathbf p');\mathbf x(-\mathbf p)),
\label{lopsym2}
\end{eqnarray} 
and from the spectral representation for the characteristic function, see Appendix A,
\begin{eqnarray}
C_{\{\lambda_n\}}(\mathbf x\mathbf pt;\mathbf x_0\mathbf p_0,t_0)=
\exp(-\beta_k(H(\mathbf x\mathbf p)-H(\mathbf x_0\mathbf p_0)))C_{\{\beta_{nk}-\lambda_n\}}(\mathbf x_0(-\mathbf p_0)t;\mathbf x(-\mathbf p)t_0).~~
\label{csym3}
\end{eqnarray} 
From the inverse transform of (\ref{charham})
\begin{eqnarray}
P(\mathbf Q,\mathbf x\mathbf pt,\mathbf x_0\mathbf p_0t_0)=
\int_{-i\infty}^{+i\infty}\prod_n\frac{d\lambda_n}{2\pi i}\exp(-\bm\lambda\mathbf Q)
C_{\bm\lambda}(\mathbf x\mathbf pt,\mathbf x_0\mathbf p_0t_0),
\label{pham}
\end{eqnarray}
we readily infer the symmetry
\begin{eqnarray}
P(\mathbf Q,\mathbf x\mathbf pt,\mathbf x_0\mathbf p_0t_0)=&&
\exp(-\beta_k(H(\mathbf x\mathbf p)-H(\mathbf x_0\mathbf p_0)))\times
\nonumber
\\
&&\exp(-\sum_n\beta_{nk}Q_n)P(-\mathbf Q,\mathbf x_0(-\mathbf p_0)t,\mathbf x(-\mathbf p)t).
\label{psymaux2}
\end{eqnarray}
We conclude that the previous analysis in the case of a single degree of freedom
can be carried over unchanged to the case of a Hamiltonian system with
many degrees of freedom.
\section{\label{sum}Summary and Conclusion}
In the present paper we have considered general properties of systems
driven by multiple heat reservoirs. The discussion has been based on 
the joint distribution for the heat transfers and system coordinates.
Analysing the Fokker-Planck equation for the joint distribution we
have identified a fundamental symmetry which relate the positive heat
transfers associated with the non equilibrium progression of the system
from an initial state to final state to the negative heat transfers associated
with the reverse transition from the final state to the initial state. This
symmetry which is specific to the multi reservoir case permits i) a 
generalisation of the asymptotic long time fluctuation theorem for the
large deviation function or, equivalently, the cumulant generating function
and yields corrections to the tails in the heat distributions, ii) a derivation of 
a recent exact fluctuation theorem for systems initially in equilibrium with
a reservoir. Extending the analysis to the time dependent case 
we have also presented iii) a derivation of the integral fluctuation theorem.
For simplicity the main analysis is based on a model with one degree of freedom
but also holds unaltered for systems with many degrees of freedom.
\section*{Appendices}
\subsection{Spectral representations}
The spectral representations of $L_{\bm\lambda}$ and $C_{\bm\lambda}$ incorporate
in a convenient form the properties of the Fokker-Planck equation (\ref{fp2})-(\ref{lop2})
\cite{Risken89,vanKampen92}. Assuming a non degenerate spectrum and
introducing a bi-orthogonal set, for details see \cite{Risken89}, i.e.,
an orthonormal and complete set of left and right eigenstates,
according to the eigenvalue equations 
\begin{eqnarray}
&&\int dx'L_{\bm\lambda}(x,x')\Psi^R_n(x',\bm\lambda)=\mu_n(\bm\lambda)\Psi^R_n(x,\bm\lambda),
\label{ap11}
\\
&&\int dx'\Psi^L_n(x',\lambda) L_\lambda(x',x)=\mu_n(\lambda)\Psi^L_n(x,\lambda),
\label{ap12}
\end{eqnarray}
with completeness and orthogonality properties
\begin{eqnarray}
&&\sum_n\Psi^R_n(x,\bm\lambda)\Psi^L_n(x',\bm\lambda)=\delta(x-x'),
\label{ap21}
\\
&&\int dx\Psi^R_n(x,\bm\lambda)\Psi^L_m(x,\bm\lambda)=\delta_{nm}.
\label{ap22}
\end{eqnarray}
Thus, setting $C_{\bm\lambda}(xt,x_0t_0)=C_{\bm\lambda}(x,t-t_0)\delta(x-x')$, we obtain the spectral forms 
\begin{eqnarray}
&&L_{\bm\lambda} (x,x')=\sum_n\mu_n(\bm\lambda)\Psi^R_n(x,\bm\lambda)\Psi^L_n(x',\bm\lambda),
\label{spec1}
\\
&&C_{\bm\lambda}(xt,x_0t_0)=\sum_n\exp(\mu_n(\bm\lambda)(t-t_0))\Psi^R_n(x,\lambda)\Psi^L_n(x_0,\bm\lambda).~~~~
\label{spec2}
\end{eqnarray}
We assume that the eigenvalue spectrum $\{\mu_n(\bm\lambda)\}$ forms a decreasing sequence
with largest eigenvalue $\mu_0(\bm\lambda)$. Moreover, completeness yields the 
following boundary condition $C_{\bm\lambda}(xt,x_0t_0)=\delta(x-x_0)$ for $t=t_0$. 

Applying the symmetry (\ref{csym}) to the spectral form (\ref{spec1}) we find  
\begin{eqnarray}
&&\mu_n(\{\nu_{nm}\})=\mu_n(\{\beta_{nm}-\nu_{nm}\}),
\label{ap31}
\\
&&\Psi^R_n(x,\{\lambda_n\})=\exp(-\beta_kU(x))\Psi^L_n(x,\{\beta_{nk}-\lambda_n\}).
\label{ap32}
\end{eqnarray}
From the leading eigenvalue $\mu_0$ we thus obtain the generalisation (\ref{cgf2})
of the usual asymptotic fluctuation theorem. We also infer the prefactor
$A_n(xx_0,\bm\lambda)=\Psi^R_n(x,\bm\lambda)\Psi^L_n(x_0,\bm\lambda)$.
\subsection{The symmetry}
In order to demonstrate the symmetry (\ref{lopsym}), i.e.,
\begin{eqnarray}
L_{\{\lambda_n\}}(x,x')=\exp(-\beta_k(U(x)-U(x')))L_{\{\beta_{nk}-\lambda_n\}}(x',x),
\label{sym1}
\end{eqnarray}
we analyse in more detail the Liouville operator
\begin{eqnarray}
L_{\bm\lambda}(x,x')=\sum_n\Gamma_n\bigg[T_n\frac{d^2}{dx^2}+U'(1-2\lambda_n T_n)\frac{d}{dx}+
(1-\lambda_n T_n)(U''-\lambda_n U'^2)\bigg]\delta (x-x').~~
\label{sym2}
\end{eqnarray}
Using the identities
\begin{eqnarray}
e^{\beta_kU(x)}\frac{d}{dx}\delta(x-x')e^{-\beta_kU(x')}&&=\Big(\frac{d}{dx}-\beta_kU'(x)\Big)\delta (x-x'),
\label{sym31}
\\
e^{\beta_kU(x)}\frac{d^2}{dx^2}\delta(x-x')e^{-\beta_kU(x')}&&=
\Big(\frac{d}{dx}-\beta_kU'(x)\Big)\Big(\frac{d}{dx}-\beta_kU'(x)\Big)\delta (x-x'),
\label{sym32}
\\
U'(x)\frac{d}{dx}\delta(x-x')&&=-\Big(U'(x')\frac{d}{dx'}-U''(x')\Big)\delta(x-x'),
\label{sym33}
\end{eqnarray}
and setting $\lambda_n\rightarrow\beta_{nk}-\lambda_n$ it is easy to show that
\begin{eqnarray}
e^{\beta_kU(x)}L_{\{\lambda_n\}}(x,x')e^{-\beta_kU(x')}=L_{\{\beta_{nk}-\lambda_n\}}(x',x),
\label{sym4}
\end{eqnarray}
and the symmetry (\ref{sym1}) follows.
\subsection{Mapping to Schr\"odinger case}
The similarity transformation
\begin{eqnarray}
g(x,\bm\lambda)=U(x)\frac{\sum_n\Gamma_n T_n(\lambda_n-\beta_n/2) }{\Gamma T},
\label{ap4}
\end{eqnarray}
maps the Liouville operator $L_{\bm\lambda}(x.,x')$ in (\ref{lop2}) to the Hermitian Schr\"odinger
form $L_{\bm\lambda}^S(x,x')$, i.e., 
\begin{eqnarray}
L^S_{\bm\lambda}(x,x')=\exp(-g(x))L_{\bm\lambda}(x,x')\exp(g(x')),
\label{ap5}
\end{eqnarray}
where 
\begin{eqnarray}
L^S_{\bm\lambda}(x)=
\Gamma\left[T\frac{d^2}{dx^2}+\frac{1}{2}U''-\frac{1}{4T}\left(\frac{\omega}{\Gamma}\right)^2U'^2\right].
\label{ap6}
\end{eqnarray} 
The associated eigenvalue problem is
\begin{eqnarray}
\int dx'L_{\bm\lambda}^S(x,x')\Psi_n(x',\bm\lambda)=\mu_n(\bm\lambda)\Psi_n(x,\bm\lambda),
\label{ap7}
\end{eqnarray}
with complete and orthogonal eigenstates,
\begin{eqnarray}
&&\sum_n\Psi_n(x,\bm\lambda)\Psi_n(x',\bm\lambda)=\delta(x-x'),
\label{ap81}
\\
&&\int dx\Psi_n(x,\bm\lambda)\Psi_m(x,\bm\lambda)=\delta_{nm}.
\label{ap82}
\end{eqnarray}
The spectral representation of $L_{\bm\lambda}(x,x')^S$ has the form
\begin{eqnarray}
L_{\bm\lambda}^S (x,x')=\sum_n\mu_n(\bm\lambda)\Psi_n(x,\bm\lambda)\Psi_n(x',\bm\lambda).
\label{ap9}
\end{eqnarray}
We conclude that the eigenvalues $\mu(\bm\lambda)$ of $L_{\bm\lambda}$
and $L_{\bm\lambda}^S$ are identical, whereas the eigenstates are transformed
according to
\begin{eqnarray}
&&\Psi^R_n(x,\bm\lambda)=\exp(g(x,\bm\lambda))\Psi_n(x,\bm\lambda),
\label{101}
\\
&&\Psi^L_n(x,\bm\lambda)=\exp(-g(x,\bm\lambda))\Psi_n(x,\bm\lambda).
\label{102}
\end{eqnarray}
We also note that the identity $g(x,\{\lambda_n\})+g(x,\{\beta_{nk}-\lambda_n\})=\beta_kU$
ensures consistency with the symmetry (\ref{ap32}).
\subsection{Harmonic potential}
For a harmonic potential $U(x)=x^2/2$ the Schr\"odinger operator takes the form
\begin{eqnarray}
L^S_{\bm\lambda}(x)=
\Gamma\left[T\frac{d^2}{dx^2}+\frac{1}{2}-\frac{1}{4T}\left(\frac{\omega}{\Gamma}\right)^2x^2\right],
\label{ap11b}
\end{eqnarray} 
corresponding to a quantum mechanical particle in the potential
$(1/4T)(\omega/\Gamma)^2x^2-1/2$. The spectrum and eigenstates are given by 
\cite{Landau59c}
\begin{eqnarray}
&&\mu_n(\bm\lambda)=\frac{1}{2}(\Gamma-\omega(\bm\lambda))-\omega(\bm\lambda) n,
\label{ap121}
\\
&&\Psi_n(x)=\Big(\frac{\omega}{2\pi\Gamma T}\Big)^{1/4}
\Big(\frac{1}{2^n n!}\Big)^{1/2}
\exp\Big(-\frac{\omega x^2}{4\Gamma T}\Big)H_n\Big(x\sqrt{\frac{\omega}{2\Gamma T}}\Big),
\label{ap122}
\end{eqnarray} 
where $H_n$ is the Hermite polynomial and $n=0,1,2,\cdots$.
\subsection{Time dependent potential} 
Here we extend our discussion of a particle in a time independent 
potential  driven by multiple reservoirs to the case where the system
is subject to a time dependent protocol modeled here by a potential
$U(xt)$ depending parametrically on $t$. More precisely, denoting 
the chosen protocol by $\lambda(t)$ we have set $U(x,\lambda(t))\equiv U(xt)$.
The equations of motion (\ref{lan1}) and (\ref{lan2}) then take the form
\begin{eqnarray}
&&\frac{dx(t)}{dt}=-\Gamma U'(xt)+\xi(t),
\label{lant1}
\\
&&\frac{dQ_n(t)}{dt}=-\Gamma_n(U'(xt))^2+U'(xt)\xi_n(t).
\label{lant2}
\end{eqnarray}
Correspondingly, the Fokker-Planck equation for the characteristic
function $C_{\bm\lambda}(x,t)$ is
\begin{eqnarray}
\frac{dC_{\bm\lambda} (xt)}{dt}= L_{\bm\lambda}(xt)C_{\bm\lambda}(xt),
\label{fpt2}
\end{eqnarray}
where
\begin{eqnarray}
L_{\bm\lambda}(xt)=\sum_n\Gamma_n&\bigg[&T_n\frac{d^2}{dx^2}+
U'(xt)(1-2\lambda_n T_n)\frac{d}{dx}+
\nonumber
\\
&&+(1-\lambda_n T_n)(U''(xt)-\lambda_n U'^2(xt))\bigg],~~~~~~~
\label{lopt2}
\end{eqnarray}
where we note the explicit time dependence of the Liouville operator engendered
by the time dependent potential.

Explicitly, the characteristic function has the form
\begin{eqnarray}
C_{\bm\lambda}(xt,x_0t_0)=\langle\exp\Big(\sum_n\lambda_n Q_n(t)\Big)\delta(x-f(t,x_0t_0))\rangle.
\label{deftc}
\end{eqnarray}
Here $f(t,x_0t_0)$ is a solution of the Langevin equation (\ref{lant1}) for a specific noise
realisation $\xi(t)$ defining a forward trajectory in configuration space from the initial configuration
$x_0$ at time $t_0$ to the final configuration  $x$ at time $t$. Note that a time dependent protocol is
acting along the trajectories. We, moreover, introduce the boundary and initial conditions
\begin{eqnarray}
&&C_{\mathbf 0}(xt,x_0t_0)=\langle\delta(x-f(t,x_0t_0))\rangle=P(xt,x_0t_0),
\label{bound}
\\
&&C_{\bm\lambda}(xt,x't)=\delta(x-x').
\label{init}
\end{eqnarray}

Since the Liouville operator has an explicit time dependence we solve (\ref{fpt2})
by iteration. Introducing for convenience the operator notation 
$\hat C_{\bm\lambda}(tt')\equiv C_{\bm\lambda}(xt,x't')$, $\hat I=\delta(x-x')$,
and $\hat L_{\bm\lambda}(t)\equiv L_{\bm\lambda}(xx',t)=L_{\bm\lambda}(xt)\delta(x-x')$
we obtain
\begin{eqnarray}
\hat C_{\bm\lambda}(tt')=\hat I+\int_{t'}^tdt''\hat L_{\bm\lambda}(t'')
+\int_{t'}^tdt''\int_{t'}^{t''}dt''' \hat L_{\bm\lambda}(t'') \hat L_{\bm\lambda}(t''')
+\cdots.
\label{exp}
\end{eqnarray}
Next, using time ordering, i.e., $T(\hat L(t)\hat L(t'))=\hat L(t)\hat L(t')$ for $t>t'$ and
$T(\hat L(t)\hat L(t'))=\hat L(t')\hat L(t)$ for $t'>t$, we have more compactly  \cite{Zinn-Justin89}
\begin{eqnarray}
\hat C_{\bm\lambda}(tt')=
T\Big[\exp\Big(\int_{t'}^t dt'' \hat L_{\bm\lambda}(t'')\Big)\Big],
\label{time1}
\end{eqnarray}
or expanded in matrix form
\begin{eqnarray}
C_{\bm\lambda}(xt,x't')=
T\Big[\exp\Big(\int_{t'}^t dt'' \hat L_{\bm\lambda}(t'')\Big)\Big](xx').
\label{time2}
\end{eqnarray}
The fundamental symmetry of the Liouville operator only refers
to the configuration coordinate $x$ and does not involve the
time dependence of the potential. Thus, we obtain as in (\ref{lopsym})
the symmetry
\begin{eqnarray}
L_{\{\lambda_n\}}(xx',t)=\exp(-\beta_k(U(xt)-U(x't)))L_{\{\beta_{nk}-\lambda_n\}}(x'x,t),
\label{sym}
\end{eqnarray}
or in operator form, introducing $\exp(\beta_k\hat U(t))\equiv \exp(\beta_kU(xt))\delta(x-x')$,
\begin{eqnarray}
L_{\{\lambda_n\}}(t)=\exp(-\beta_k\hat U(t))\hat L_{\{\beta_{nk}-\lambda_n\}}^\ast (t)\exp(\beta_k\hat U(t)),
\label{symop}
\end{eqnarray}
where $\ast$ indicates the transposed (hermitian conjugate) operator.
\newpage
%

\end{document}